\documentclass[%
prl,reprint,
superscriptaddress,
nofootinbib,
amsmath,amssymb,
aps,
]{revtex4-1}

\usepackage{graphicx}
\usepackage{dcolumn}
\usepackage{bm}
\usepackage{textcomp}
\usepackage[pdftex,colorlinks=true,bookmarks=false,citecolor=blue,urlcolor=blue]{hyperref}
\usepackage{cancel}
\newcolumntype{?}{!{\vrule width 1.1pt}}

\usepackage[toc,page]{appendix}

\allowdisplaybreaks
\newcommand{\mrm}[1]{\mathrm{#1}}

\newcommand{\Avec}{\mathbf{A}}

\newcommand{\Svec}{\mathbf{S}}	
\newcommand{\Evec}{\mathbf{E}}	
\newcommand{\Bvec}{\mathbf{B}}	
\newcommand{\Jvec}{\mathbf{J}}	

\newcommand{\pvec}{\mathbf{p}}	

\newcommand{\vvec}{\mathbf{v}}  
\newcommand{\evec}{\mathbf{e}}	

\DeclareSymbolFont{lettersA}{U}{txmia}{m}{it}
\DeclareMathSymbol{\real}{\mathord}{lettersA}{"92} 
\DeclareMathSymbol{\cplx}{\mathord}{lettersA}{"83} 

\newcommand{\EP}{electromagnetic pulse}
\newcommand{\EPs}{electromagnetic pulses}
\newcommand{\EB}{electron beam}
\newcommand{\EBs}{electron beams}

\newcommand{\ITedit}[1]{{\color{black}#1}}

\newcommand{\rev}[1]{{\color{black}#1}}
\newcommand{\revES}[1]{{\color{black}#1}}
\newcommand{\revIT}[1]{{\color{black}#1}}

\hypersetup{
   pdftitle=Electron beam driven generation of sub-cycle pulses
}

\begin{document}


\title{Electron beam driven generation of \rev{frequency-tunable \\ isolated} relativistic sub-cycle pulses}

\author{I.~Thiele}
\email{illia-thiele@web.de}
\affiliation{Department of Physics, Chalmers University of Technology, SE-412 96 G{\"o}teborg, Sweden}
\author{E.~Siminos}
\affiliation{Department of Physics, University of Gothenburg, SE-412 96 G{\"o}teborg, Sweden}
\author{T.~F{\"u}l{\"o}p}
\affiliation{Department of Physics, Chalmers University of Technology, SE-412 96 G{\"o}teborg, Sweden}

\date{\today}

\begin{abstract}
	We propose a novel scheme \rev{for frequency-tunable} sub-cycle electromagnetic pulse generation. 
	To this end a pump electron beam is injected into an electromagnetic seed pulse as the latter is reflected by a mirror. The electron beam is shown to be able to amplify the field of the seed pulse while upshifting its central frequency and reducing its number of cycles. 
	We demonstrate the amplification by means of 1D and 2D particle-in-cell simulations. In order to explain and optimize the process, a model based on fluid theory is proposed. 
	We estimate that using currently available electron beams and terahertz pulse sources, our scheme is able to produce mJ-strong mid-infrared sub-cycle pulses.		
\end{abstract}

\maketitle



Generation of few cycle electromagnetic pulses has steadily advanced, driven by applications which require probing or control of ultra-fast processes~\cite{Corkum07,Krausz14}. Recently a lot of effort has been devoted to
producing sub-cycle pulses in which the time-envelope is modulated at time scale shorter than a single cycle. Such pulses bring temporal resolution to its ultimate limits and are unique tools for the control of electron motion in solids~\cite{Hohenleutner15}, electron tunneling in nano-devices~\cite{Rybka16}, reaction dynamics at the
electronic level~\cite{Kling06}, as well as the generation of isolated attosecond and zeptosecond X-ray pulses~\cite{PhysRevLett.111.033002}. 
Several methods like optical synthesis or parametric amplification have been developed for the generation of sub-cycle pulses from the THz to X-ray regimes~(see the review \cite{Manzoni15}). 
\rev{While for few-cycle pulse durations these methods can lead to mJ pulse energies, the energies of sub-cycle pulses are limited to a few \textmu J}.
The main limitation of \ITedit{the typically used} parametric amplification methods is the material damage threshold under intense fields~\cite{Rivas17}.
On the other hand, methods exploiting plasmas or electron beams as a frequency conversion medium, such as high-harmonic generation from solid targets~\cite{teubner2009}, Thomson scattering amplification~\cite{esarey1993}, scattering by relativistic mirrors~\cite{Bulanov16} and frequency down-conversion in a plasma wake~\cite{Tsung29,nie2018} are not subject to a damage threshold. However, these methods are not able to generate isolated sub-cycle pulses. 

\begin{figure}[b]
	\centering
	\includegraphics[width=0.9\columnwidth]{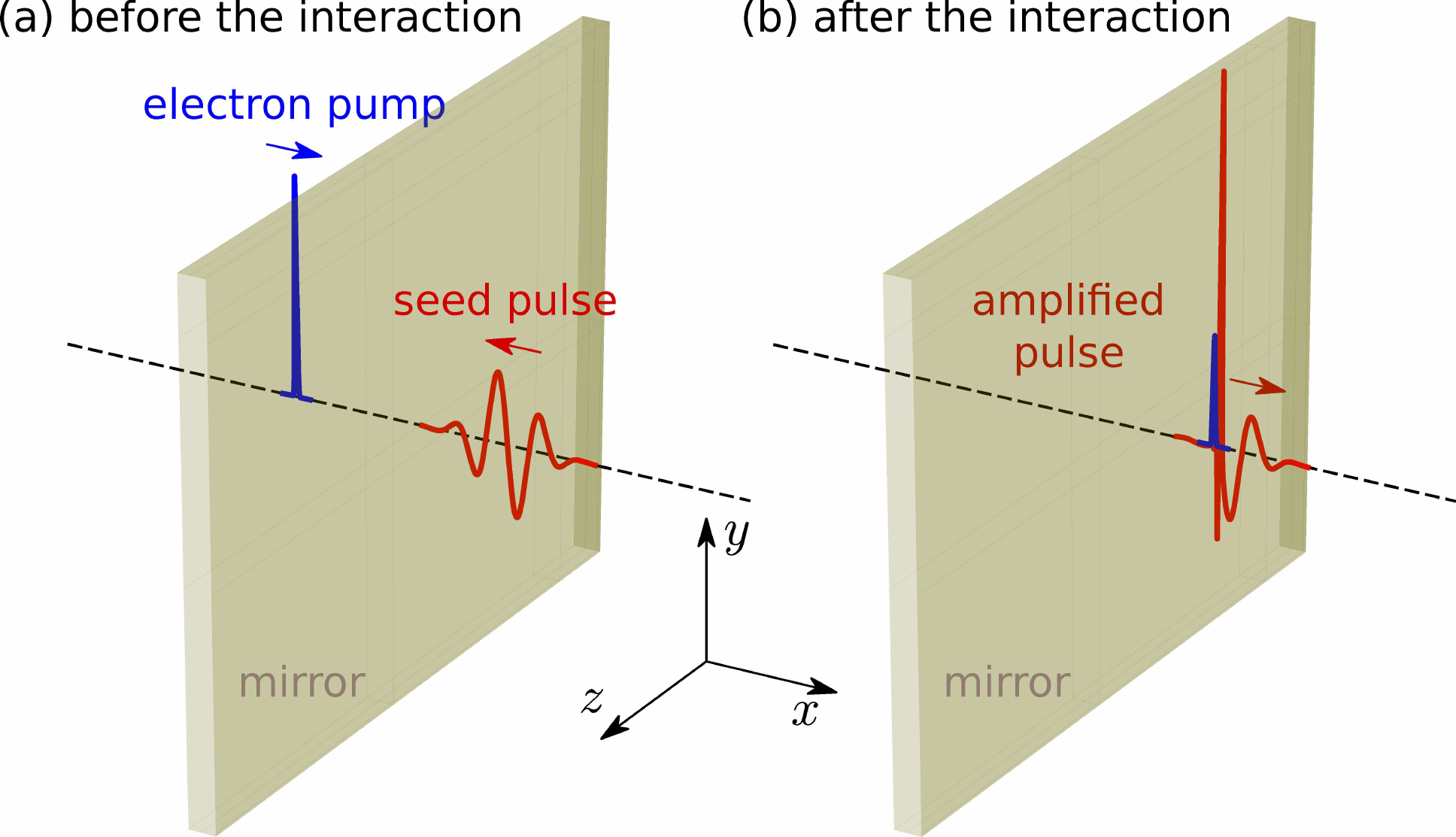}
	\caption{Schematic representation of the electron beam driven amplification~(EBDA) scheme:~(a)~The counter-propagating seed electromagnetic pulse and pump electron beam are moving towards a mirror (thin foil).~(b) The \EP\ is reflected by the mirror and interacts with the \EB\ {as it exits through the mirror}, leading to {the generation and amplification of} an intense sub-cycle pulse.}
	\label{fig:scheme}
\end{figure}

\begin{figure}
	\centering
	\includegraphics[width=1.0\columnwidth]{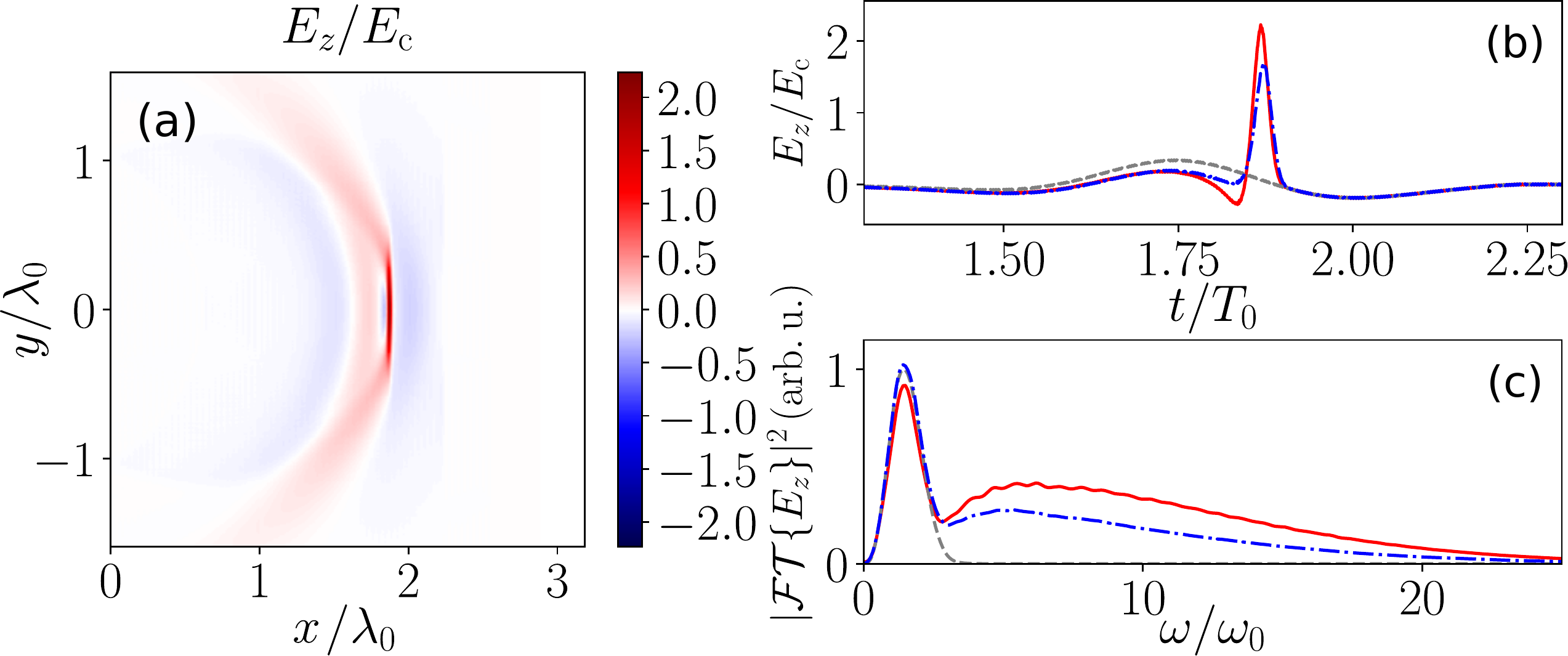}
	\includegraphics[width=1.0\columnwidth]{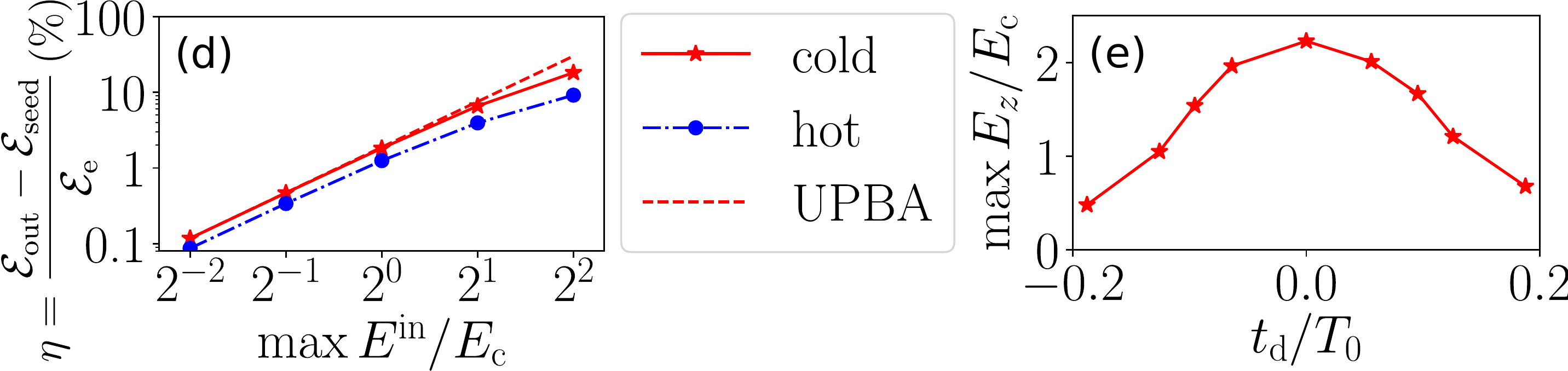}
	\caption{(a)~Electric field snapshot after the interaction of a strongly focused low-frequency pulse with a {mono-energetic} electron beam passing the standing mirror at $x=0$. Corresponding on-axis signal~(b) and spectrum~(c) demonstrating the generation of an intense higher-frequency sub-cycle pulse
		{with a cold mono-energetic electron beam~(solid red lines), a beam with 150\% energy spread~having the same total energy (dash-dotted blue lines) and without the electron beam~(dashed gray lines).}
		{(d)~Conversion efficiency versus seed field strength. (e)~Amplified peak electric field dependence on delay time $t_\mrm{d}$.
		} 
		The parameters are: $t_\mrm{e}=0.016T_0$, {$y_\mrm{e}=0.3\lambda_0$}, $n_\mrm{e}^\mrm{max}=28.3n_\mrm{c}$, $\gamma_\mrm{e}=20$, $t_0=0.21T_0$ and $E_0^\mrm{in}=E_\mrm{c}$~{[except in (d)]}. The simulation was performed with 1600 points per \EP\ carrier wavelength along $x$, 100 points along $y$, 1608 points per \EP\ carrier oscillation and {100 macroparticles} per cell. 
		\revES{We consider an aluminium foil of thickness $1.4c/\omega_0$ acting as a mirror. Since such a foil does not stop MeV-electrons~\cite[p.~376]{Heitler} and ensures the reflection of the low-frequency seed pulse, we simply model it as a dense electron plasma}.}
	\label{fig:2D}
\end{figure}

\rev{In this letter, we propose a method to generate frequency-tunable isolated sub-cycle pulses reaching relativistic intensities.
We particularly focus on the mid-infrared~(MIR) regime~\cite{Liang17}.
Such pulses would lead to an ultra-strong light-matter coupling and might enable the switching of light-matter interaction within less than one cycle of light for the observation of new quantum mechanical non-adiabatic phenomena~\cite{Gunter09} or high harmonic and isolated zeptosecond pulse generation with a significantly extended frequency cut-off~\cite{PhysRevLett.111.033002,Hohenleutner15}.}
As visualized in Fig.~\ref{fig:scheme}, our scheme involves the interaction of a seed electromagnetic pulse with a short duration pump electron beam at a thin foil. The thin foil acts as a mirror reflecting the seed pulse, while the \EB\ enters in the middle of the pulse and leads to its amplification in a co-propagating configuration. 
\revIT{As will be shown below, a substantial part of the \EB\ energy can be transferred to the \EP, more than doubling its energy.
\revES{Currently available} single-cycle THz sources reaching mJ-pulse energies for central frequencies up to $\nu_\mrm{THz}=4$~THz can be employed \revES{to produce suitable seed pulses}~\cite{Vicario14,Liao18}. 
To obtain sub-cycle pulses of comparable energy {and with the central frequency in the MIR}, 10-MeV nC electron bunches which are shorter than a single THz oscillation can be produced by compact laser-wakefield accelerators (LWFA)~\cite{Goers2015,Salehi:17}.}

We demonstrate the scheme through a 2D particle-in-cell~(PIC) simulation with the code SMILEI~\cite{SMILEI}. A linearly polarized single-cycle seed pulse is focused strongly onto a thin almost perfectly reflecting foil. 
However, as will be clarified later on, our scheme can also operate with many-cycle seed pulses.
{The incoming seed pulse is focused at the mirror to obtain the $y$-polarized electric field~$\Evec=E_0^\mrm{in}\exp(-y^2/y_0^2)\exp(-t^2/T_0^2)\sin(\omega_0 t)\evec_y$, with the field amplitude $E_0^\mrm{in}=E_\mrm{c}$ where $E_\mrm{c}=cm_\mrm{e}\omega_0/q_\mrm{e}$, $\omega_0=2\pi/T_0$ is the carrier frequency corresponding to the wavelength $\lambda_0=cT_0$, $y_0=0.3\lambda_0$ characterizes the beam width, $t_0=0.21T_0$ gives the time duration and $\evec_y$ is the unit vector along $y$.}
The electron beam is entering from the back side of the foil. It is initialized with a constant gamma factor $\gamma_\mrm{e}=20$ and a Gaussian density profile {$n_\mrm{e} = n_\mrm{e}^\mrm{max}\exp(-y^2/y_\mrm{e}^2)\exp[-(x+\lambda_0)^2/(ct_\mrm{e})^2]$ with thickness} $y_\mrm{e}=0.3\lambda_0$, duration $t_\mrm{e}=0.016T_0$ and peak density $n_\mrm{e}^\mrm{max}=28.3n_\mrm{c}$, where $n_\mrm{c}=m_\mrm{e}\epsilon_0\omega_0^2/q_\mrm{e}^2$ is the critical density for a resting plasma. A snapshot of the electric field after the amplification process has been completed is presented in Fig.~\ref{fig:2D}(a). We observe a strong sub-cycle pulse around \ITedit{$x=1.91\lambda_0$}. It is well collimated compared to the residual driving \EP\ which diffracts strongly due to the tight focusing. This is an advantageous property of the scheme because of the natural separation between the seed and amplified \EP. As the on-axis electric field time-trace in Fig.~\ref{fig:2D}(b) demonstrates, already after {propagation for} two seed-wavelengths, the pulses are almost separated. The corresponding frequency spectrum in Fig.~\ref{fig:2D}(c) shows that the generated sub-cycle pulse is up-shifted by {a factor of} seven in terms of peak frequency and is therefore diffracting much less than the seed pulse. The energy of the pulse is amplified by a factor of 2.4. 

\begin{figure}[b]
	\centering
	\includegraphics[width=1.0\columnwidth]{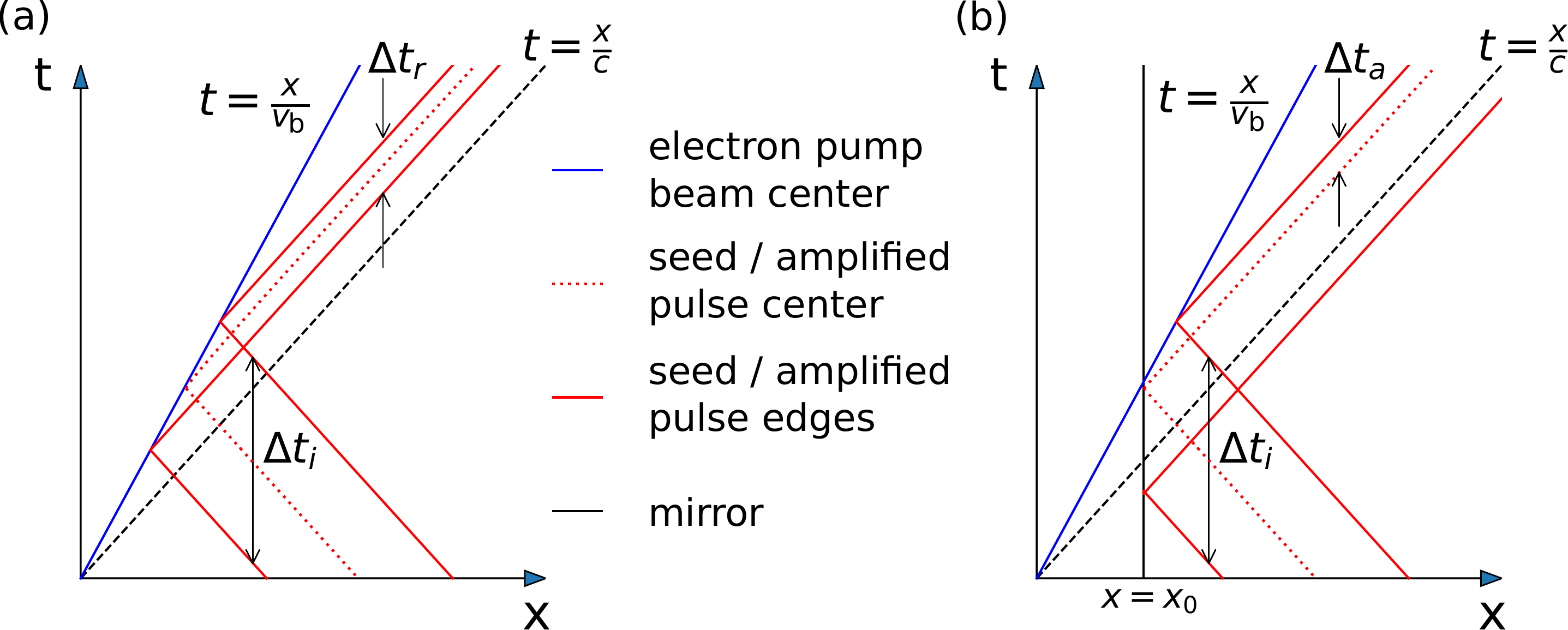}
	\caption{Space-time diagrams visualizing the \EB~(blue line) moving with  speed $v_{\mathrm b}$ and \EP~(red lines) without~(a) or with~(b) the standing mirror at $x=x_0$~[black line in (b)]: Without the mirror the whole \EP\ and with the mirror only {part} of the \EP\ interacts with the \EB\ allowing for sub-cycle pulse generation.}
	\label{fig:scheme_1C}
\end{figure}

{The process remains effective when using an electron beam with 150\% energy spread, {\emph{i.e.} with a Maxwellian-like spectrum similar to those produced by LWFA operating in the self-modulated regime~\cite{Goers2015,Salehi:17}}, see Figs.~\ref{fig:2D}(b,c). 
{{For cold electron beams} the overall efficiency $\eta=(\mathcal{E}_\mrm{out}-\mathcal{E}_\mrm{seed})/\mathcal{E}_\mrm{e}$, where $\mathcal{E}_\mrm{out}$, $\mathcal{E}_\mrm{seed}$ and $\mathcal{E}_\mrm{e}$ are the outgoing, seed pulse and electron beam energies, {ranges from $1\%$ for weak seed pulses, up to} $18\%$ for stronger seed pulses~[see Fig.~\ref{fig:2D}(d)].} {For Maxwellian-like {electron} beam spectra, the conversion efficiency is slightly lower, yet remains above $9\%$ for strong seed pulses~[see Fig.~\ref{fig:2D}(d)], implying that such electron beams are still usable for the production of mJ-level mid-IR sub-cycle pulses.}
{Moreover, amplification} is robust with respect to jitter effects~[see Fig.~\ref{fig:2D}(e)]}.
{We note that {the radiation reported here} is distinct from transition radiation~\cite{Jackson,Schroeder2004}, which can dominate for weak seed pulses but has very different properties~(see Appendices).

{In order to illustrate why a standing mirror is required in addition to the \EB\ in order to produce \emph{sub-cycle} pulses, we consider the simplified space-time diagrams in Fig.~\ref{fig:scheme_1C}. We restrict attention to 1D geometry and consider the limit of an infinitely dense and sharply rising electron beam front. Without the standing mirror~[Fig.~\ref{fig:scheme_1C}(a)] the setup is known as the relativistic flying mirror concept~\cite{Bulanov16}.} 
The solid red line indicates the edges of the incoming \EP\ which is perfectly reflected by the \EB. {Due to the double Doppler shift effect the frequency of the reflected pulse is upshifted and its amplitude amplified by a factor $\approx4\gamma_\mrm{e}^2$~\cite{Landecker52}. While the duration of the reflected pulse is shortened by the same factor, the number of cycles remains invariant.} {By contrast, {when} the standing mirror {is introduced}~[solid black line at $x=x_0$ in Fig.~\ref{fig:scheme_1C}(b)], the leading part of the \EP\ is simply reflected and not amplified. Only the trailing part interacts with the \EB\ and, thus, the number of \emph{amplified} cycles is reduced.}


We \rev{now present a simplified 1D~($\partial_z=\partial_y=0$) fluid model of the interaction (for details see Appendices) in order to illuminate the mechanism of the electron-beam-driven amplification~(EBDA).}
{The transverse fluid momentum $\pvec_\perp$ evolves according to
\begin{equation}\label{eq:dpperp}
	\partial_t\pvec_\perp=q_\mrm{e}\Evec_\perp\,\mrm{,} 
\end{equation}
where $\Evec_\perp$ is the transverse  electric field, and \ITedit{$q_\mrm{e}$ is the electron charge}.
This leads to conservation of transverse canonical momentum, {$\pvec_\perp=-q_\mrm{e}\mathbf{A}_\perp$},
where {$\Evec_\perp=-\partial_t \Avec_\perp$ and $\Avec_\perp$} is the vector potential (in the Coulomb gauge).
The transverse current reads 
\begin{equation}\label{eq:J}
  \Jvec_\perp=(q_\mrm{e}/m_\mrm{e})\pvec_\perp n_\mrm{e}/\gamma_\mrm{e}\,\mrm{,} 
\end{equation} 
where $\gamma_\mrm{e}=\sqrt{1+|\pvec|^2/(m_\mrm{e}c)^2}$, \ITedit{$m_\mrm{e}$} is the electron mass and $c$ the speed of light in vacuum.
}
If we choose a sufficiently weak seed pulse, the longitudinal momentum of the electrons dominates and $\gamma_\mrm{e} \approx p_x/(m_\mrm{e}c)$.
{This allows us to neglect the effect of the seed pulse on the longitudinal electron beam momentum, i.e., to employ an undepleted pump beam approximation~(UPBA).}
{Equation~(\ref{eq:dpperp}) is then solved together with Maxwell's equations with a source term given by Eq.~(\ref{eq:J})} for a given electron beam dynamics {with prescribed} $n_\mrm{e}(x,t)$ and $v_x(x,t) = p_x(x,t)/[m_\mrm{e}\gamma_\mrm{e}(x,t)]$. For simplicity we assume a density profile moving with a constant speed $v_\mrm{b}$ \rev{and $\gamma_\mrm{e}$}, 
\begin{equation}
	n_\mrm{e}(x,t) = n_\mrm{e}^\mrm{max}\exp\left[-(t-t_\mrm{d}-x/v_\mrm{b}){^2}/t_\mrm{e}^2\right]\,\mrm{,}
\end{equation} 
with some delay $t_\mrm{d}$. 
\ITedit{The seed pulse arriving from $x=+\infty$ is perfectly reflected by the mirror at $x=0$, i.e., the electric field at the mirror is zero. The fields can be decomposed into forward and backward propagating parts $E_z^-$ and $E_z^+$ respectively such that the seed pulse electric field at the mirror is defined by}
\begin{equation}
\ITedit{E_z^\pm(t) = \mp E_z^\mrm{max}\sin\left(\omega_0t\right)\exp\left(-t\rev{^2}/t_0^2\right)\,\mrm{.}}
\end{equation}
\ITedit{In Fig.~\ref{fig:1D_1C}(a), an example of an $n_\mrm{e}/\gamma_\mrm{e}$-profile~(dotted line) and an outgoing seed electric field~(dashed line) is shown.}

{The solution of the fluid model for a weak seed pulse and a density profile shorter than the cycle duration is shown in Fig.~\ref{fig:1D_1C}(b). We indeed observe a partial} amplification of the incoming seed pulse~(dashed line), {leading to the formation of a sub-cycle pulse in the center of the original pulse}. After the interaction, the pulse energy increases by a factor of 10 and the maximum electric field of the \EP\ is enhanced by about a factor of 14~(solid line). We compare the result of the fluid model with PIC simulations~[dark red solid line in Fig.~\ref{fig:1D_1C}(b) {and Figs.~\ref{fig:2D}(d)}] to find an excellent agreement which justifies the use of the fluid picture and the UPBA. As Fig.~\ref{fig:1D_1C}(c) shows, the spectrum of the reflected pulse is {up-shifted by a factor of \ITedit{seven}} and strongly broadened. 

{From Poynting's theorem and Eqs.~(\ref{eq:dpperp}) and (\ref{eq:J}) we can compute the energy density $U_\mrm{gain}$ transferred to the electromagnetic field at any given point in space during the interaction as~(see Appendices)
\begin{align}
U_\mrm{gain}(t)=\frac{q_e^2}{m_\mrm{e}}\int\limits_{-\infty}^t\left|\Avec_\perp\right|^2\frac{\partial}{\partial\tau}\left(\frac{n_\mrm{e}}{\gamma_\mrm{e}}\right)\,d\tau\,\mrm{.}\label{eq:U}
\end{align}
{Here we ignored terms that identically vanish after the end of the interaction.}
It is {important to note that the sign of $U_\mrm{gain}$ at any given time only depends on the rate of change of} $n_\mrm{e}/\gamma_\mrm{e}$. 
For a constant $\gamma_\mrm{e}$, the rising {part} of the electron beam gives a gain, while the descending {part} of the electron beam gives a loss. {Assuming a symmetric electron beam profile, a net gain after the end of the interaction ($t\rightarrow\infty$) requires an asymmetry in the {amplitude of} electromagnetic field vector potential $|\Avec_\perp|$.}  \rev{For a quantitative assessment $\Avec_\perp$ has to be determined by solving the full problem.} 

\begin{figure}
	\centering
	\includegraphics[width=1.0\columnwidth]{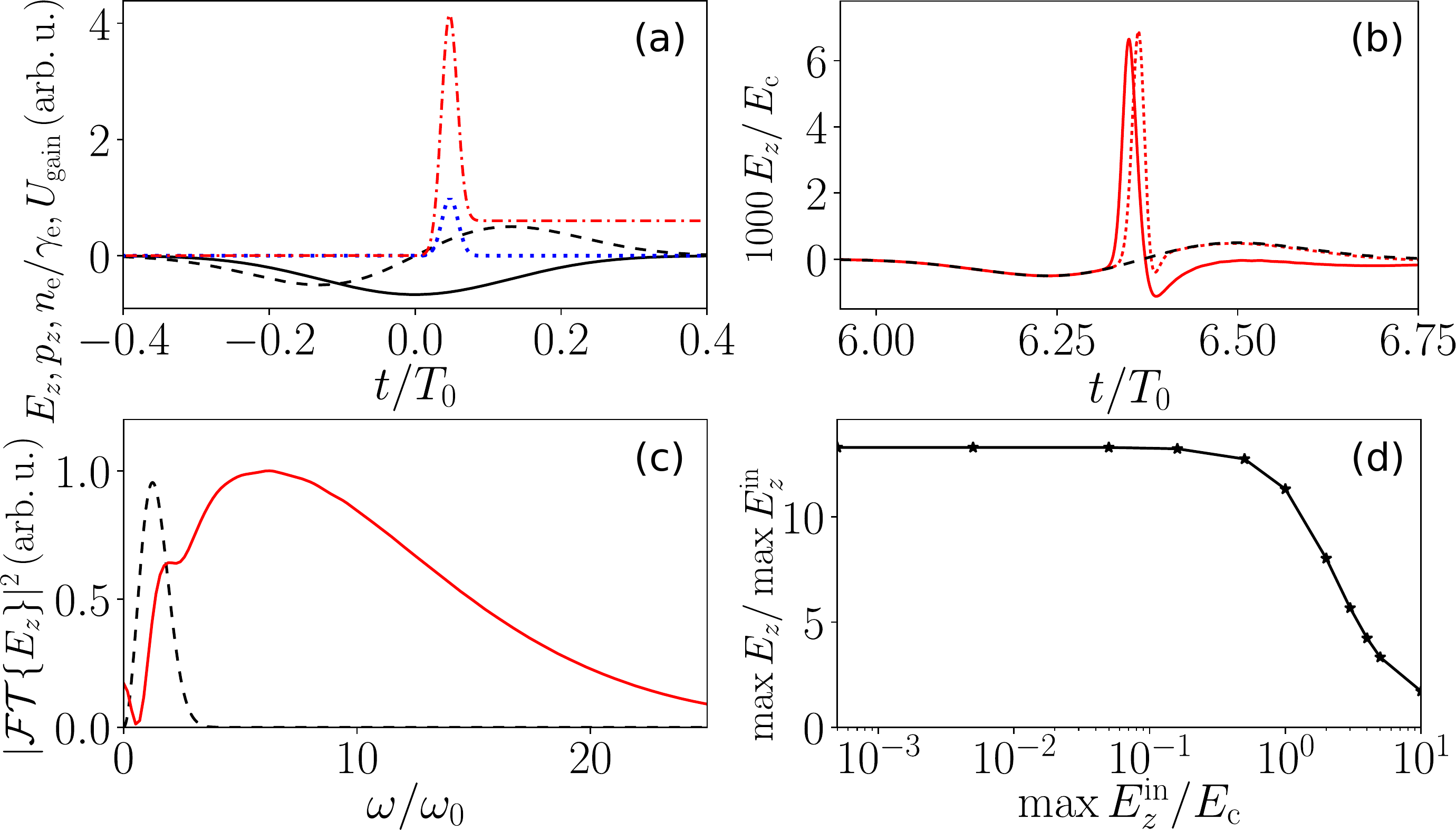}
	\caption{
		\ITedit{(a)~Visualization of the amplification mechanism: electric field approximated by the unperturbed reflected seed pulse field~(dashed line), corresponding vector potential~(solid line), electron density~(dotted line) and time-dependent energy density gain $U_\mrm{gain}$~(dash-dotted line).
		(b)~Electric field after the interaction of the incoming plane wave single-cycle pulse~(dashed line) with the electron beam according to the fluid model~(dotted line) and PIC~(solid line). (c)~Spectra according to PIC simulation. The parameters are: $t_\mrm{e}=0.016T_0$, \rev{$t_\mrm{d}=0$}, $n_\mrm{e}^\mrm{max}=28.3n_\mrm{c}$, $\gamma_\mrm{e}=20$, $t_0=0.21T_0$, $E_0^\mrm{in}=0.001E_\mrm{c}$. The distance from the mirror to the detector is $6.37c/\omega_0$. (d)~Peak electric field amplitude amplification for different incoming seed pulse amplitudes according to PIC simulations. {The remaining parameters are the same as above.}}}
	\label{fig:1D_1C}
\end{figure}

A great advantage of our scheme ({see} Fig.~\ref{fig:scheme}) is that the introduction of the standing mirror allows the electron beam to be injected into the seed pulse in a way that {such an asymmetry, and thus a net gain, can be achieved}. 
This is illustrated in Fig.~\ref{fig:1D_1C}(a) where, in order to explain the interaction in simple terms, we assumed the electric field $E_z$ to be the one of the unperturbed reflected seed pulse~(dashed line). The corresponding transverse vector potential $\mathbf{A}_\perp=-\pvec_\perp/q_\mrm{e}$ for our example can be then directly computed from Eq.~(\ref{eq:dpperp}) and is also presented in Fig.~\ref{fig:1D_1C}(a)~(solid line). This gives the evolution of $U_\mrm{gain}$~(dash-dotted line), which first increases at the rising edge of $n_\mrm{e}/\gamma_\mrm{e}$~[dotted line in Fig.~\ref{fig:1D_1C}(a)] and then decreases at the descending edge. \rev{Since in our example we have chosen a slightly positive electron beam delay $t_\mrm{d}=0.05T_0$, a final local nonzero energy gain can be expected. We shall note that injecting the beam with $t_\mrm{d}=0$ as for Figs.~\ref{fig:1D_1C}(b-c) leads also to an energy gain after some propagation since the electron beam is slower than the seed pulse. \revES{In order to take into account other effects that may become important, such as} the modification of the electric field due to its amplification \revES{or the contribution of} the incoming part of the seed pulse, the full model needs to be solved.} 

{For \EB\ duration that is much shorter than the laser cycle we may Taylor-expand $\Avec_\perp(x,t)$ in Eq.~(\ref{eq:U}) around $t=t_d$} 
\begin{equation}\label{eq:UgainShortBunch}
  U_\mrm{gain}(+\infty)=-\frac{q_e^2}{m_\mrm{e}}\frac{\partial}{\partial t}|\Avec_\perp{(t_\mrm{d})}|^2\int_{-\infty}^{+\infty} n_\mrm{e}/\gamma_\mrm{e}\,dt\,. 
\end{equation}
{This shows explicitly the dependence of the energy gain on the electromagnetic field profile when the bunch exits through the mirror.} 
{We see that, in this limit, the maximum energy gain is independent} of the \EB\ duration for a constant charge. 
\rev{Equation~(\ref{eq:U}) can be used to predict many other trends, such as a decrease of the amplification for longer \EBs, the possibility to maintain the amplification using many-cycle seed pulses if $t_\mrm{d}\neq0$ or using a chirp~(see Appendices for details)}.


The fluid model, within the UPBA, predicts a linear increase of the amplification with the amplitude of the seed pulse. \rev{As Figs.~\ref{fig:2D}(d), \ref{fig:1D_1C}(d) show, this is true up to relativistic seed pulse amplitudes.} 
This feature of our scheme implies the possibility to up-scale the amplitudes of the sub-cycle pulses up to relativistic intensities. 


\begin{figure}[b]
	\centering
	\includegraphics[width=1.0\columnwidth]{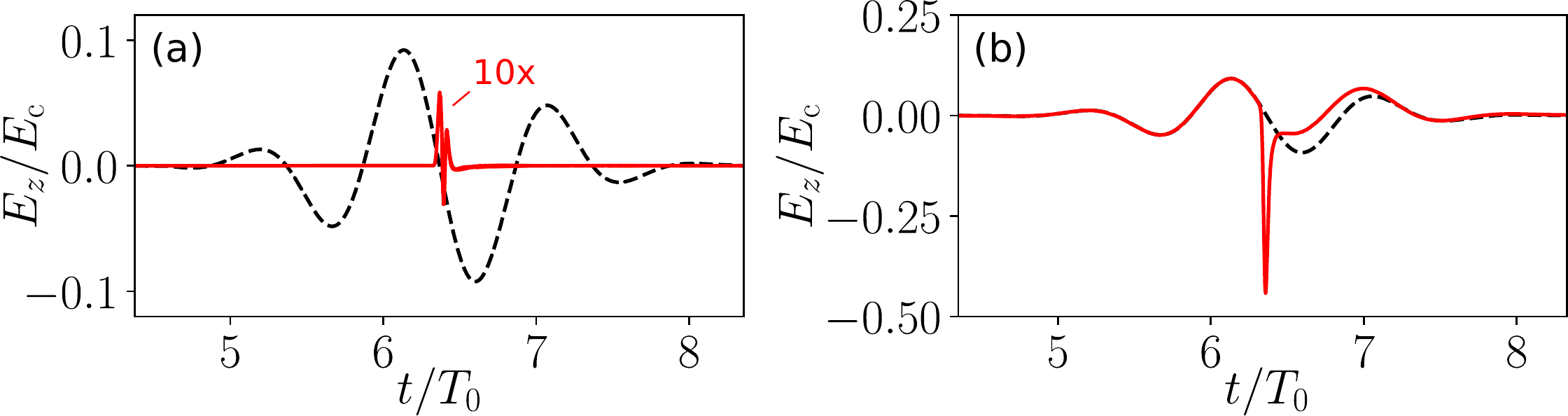}
	\caption{Electric field time-traces before~(dashed lines) and after~(solid lines) the interaction using a few-cycle seed pulse and an underdense \EB\ without~(a) or with~(b) the mirror.  
		The field after the interaction without the mirror [solid line in (a)] has been magnified by a factor of 10 for a better comparison.	
		The parameters are: \ITedit{$t_\mrm{e}=0.016T_0$, $n_\mrm{e}^\mrm{max}=5.66n_\mrm{c}$, $\gamma_\mrm{e}=10$, $t_0=0.85T_0$, $E_0^\mrm{in}=0.1E_\mrm{c}$}.
		{The simulations were performed with 1600 points per \EP\ carrier wavelength and with 1000 particles per cell. The mirror has been modeled as a dense electron plasma with thickness \ITedit{$0.22\lambda_0$}.}	
	}
	\label{fig:1D_2C}
\end{figure}

{As the example in Fig.~\ref{fig:1D_2C} demonstrates, our sub-cycle-pulse generation and amplification scheme works} not only for single-cycle but also for few-cycle driving \EPs. Actually, the scheme would lead to amplification for any duration of the driving pulse. As long as the \EB\ duration is small compared to \EP\ cycle duration $2\pi/\omega_0$, {Eq.~(\ref{eq:UgainShortBunch})} predicts a potential energy gain independent on the number of cycles in $\Avec_\perp$. 

{The cases demonstrated in Fig.~\ref{fig:1D_1C} and Fig.~\ref{fig:1D_2C} also differ in terms of the maximum value for $n_\mrm{e}/\gamma_\mrm{e}$. In the former case, the electron beam is overdense~($n_\mrm{e}^\mrm{max}/\gamma_\mrm{e}>n_\mrm{c}$) for the carrier frequency $\omega_0$, while in the latter case it is underdense~($n_\mrm{e}^\mrm{max}/\gamma_\mrm{e}<n_\mrm{c}$), i.e., the reflectivity due to the electron beam itself is almost zero. Nevertheless, due to the mirror the incident \EP\ is fully reflected. Thus, its amplitude is amplified by a factor of 4.8 and the energy is doubled even in the underdense case. By contrast, without the mirror, the reflected pulse contains only 1.7\% of the initial \EP\ energy, its amplitude is diminished by 20 times compared to the seed pulse and no sub-cycle pulse is generated.}

Up to now we presented our results in normalized units: in particular frequencies were normalized to $\omega_0$, durations to $T_0=2\pi/\omega_0$, electric fields to $E_\mrm{c}=cm_\mrm{e}\omega_0/q_\mrm{e}$ and densities to the critical density $n_\mrm{c}$. This implies the possibility to {tune the frequency spectrum of the generated sub-cycle pulse with the input parameters.}
{The output frequencies can be tuned proportionally with the seed carrier frequency $\omega_0$ if $n_\mrm{e}/\gamma_\mrm{e}$ is increased with $\omega_0^2$, the \EB\ duration and transverse size are decreased with $1/\omega_0$.}
{This corresponds to a reduction of the beam charge and sub-cycle pulse energy with $1/\omega_0$, but, a peak electric field amplitude rise with $\omega_0$.}

For the cases we are looking at in this Letter, the central frequency up-shifts by about a factor of 10. This leads to the frequency conversion key as presented in Table~\ref{tab} {including the necessary bunch duration, transverse size and charge computed from the electron density while assuming equal size in both transverse dimensions.}


\begin{table*}
	\begin{tabular}{ c c || c | c | c || c c }
		Seed {carrier} & frequency & {Charge [pC]} & {Transverse size [$\mu$m]} & {Bunch duration [fs]} & Output  & {central} wavelength \\ \hline
		THz      & $1-10$~THz & {500-5000} & {30-300} & {10-100} & (Mid)-IR & $3-30$~\textmu m\\ 
		(Mid)-IR & $10-100$~THz & {50-500} & {3-30} & {1-10} & Optical & $300$~nm $-\,3$~\textmu m\\  
		Optical  & $100-1000$~THz & {5-50} & {0.3-3} & {0.1-1} & EUV & $30$~nm $-\,300$~nm  
	\end{tabular}
	\caption{Frequency conversion key between seed and amplified \EP\ for the sub-cycle generation mechanism.} 
	\label{tab}
\end{table*}

In summary, we have proposed a scheme for the generation of isolated, intense,  sub-cycle pulses which is based on the interaction of an electron beam with a seed electromagnetic pulse reflected by a mirror. The mirror is a crucial element which allows to introduce the electron beam with the correct phase into the fully reflected seed pulse.
This ensures an efficient energy conversion from the beam to the pulse leading up to relativistic intensities and down to sub-cycle duration. In particular, we have shown that using currently available intense terahertz pulse sources and laser-wakefield-accelerated electron beams, mJ-strong mid-infrared sub-cycle pulses can be generated. We believe that our proposed scheme will trigger further theoretical and experimental investigations of both, intense sub-cycle pulse sources and applications.

\section*{Acknowledgments}
  The authors thank M.~Grech for helpful discussions and the anonymous referees for helpful comments.
  This work was supported by the Knut and Alice Wallenberg
  Foundation, the European Research Council (ERC-2014-CoG grant
  647121) and by the Swedish Research Council, Grant No. 2016-05012. 
  Numerical simulations were performed using computing resources at Grand {\'E}quipement National pour le Calcul Intensif (GENCI, Grants No.~A0030506129 and No.~A0040507594) and Chalmers Centre for Computational Science and Engineering (C3SE) provided by the Swedish National Infrastructure
  for Computing (SNIC, Grant SNIC 2017/1-484, SNIC 2017/1-393, SNIC 2018/1-43).

\begin{appendices}

\section{Cold-fluid theory of electron-beam-driven amplification in 1D}
\label{app:EBDA}

In the main article, we investigate the interaction of a seed electromagnetic pulse with an electron beam passing {through} a conducting foil. As has been shown by 1D and 2D particle-in-cell simulations, this interaction process leads to generation of an intense sub-cycle pulse. We call this process electron-beam-driven amplification~(EBDA). In the following, we present the model which is used in the main article to explain the amplification process.

We assume that the dynamics of the electron beam follows the Euler equation for a cold fluid
\begin{equation}
\partial_t\pvec+(\vvec\cdot\nabla)\pvec = q_\mrm{e}(\Evec +  \vvec\times\Bvec)\label{eq:Euler}\mrm{,}
\end{equation}
where $\pvec$ is the electron fluid momentum, $\vvec$ is the electron fluid velocity, $q_\mrm{e}$ is the electron charge, $\Evec$ is the electric field and $\Bvec$ is the magnetic field.
The longitudinal component of this equation in 1D~(assuming $\partial_y=0=\partial_z$) writes
\begin{equation}
\partial_t p_x + v_x\partial_x p_x = q_\mrm{e}\left(E_x + v_y B_z - v_z B_y\right)\mrm{.}\label{eq:p_x}
\end{equation}
The transverse components, without the presence of a longitudinal magnetic field~($B_x=0$), read 
\begin{align}
	&\partial_tp_y+v_x\partial_x p_y = q_\mrm{e}(E_y -  v_x B_z)\label{eq:p_y}\\
	&\partial_tp_z+v_x\partial_x p_z = q_\mrm{e}(E_z +  v_x B_y)\label{eq:p_z}\mrm{.}
\end{align}
We can introduce the vector potential in the Coulomb gauge $\Avec$ with the transverse components $A_y$ and $A_z$ by
\begin{align}
	\Evec = - \partial_t \Avec\mrm{,}\label{eq:Gauge}\qquad
	\Bvec = \nabla\times\Avec\mrm{.}
\end{align}
Then, Eqs.~(\ref{eq:p_y}),~(\ref{eq:p_z}) can be rewritten to
\begin{align}
	&\partial_t(p_y+q_\mrm{e}A_y) = -v_x\partial_x(p_y +  q_\mrm{e} A_y)\label{eq:p_y_2}\\
	&\partial_t(p_z+q_\mrm{e}A_z) = -v_x\partial_x(p_z +  q_\mrm{e} A_z)\mrm{.}\label{eq:p_z_2}
\end{align}
Initially, before the electron bunch interacts with the seed pulse 
\begin{align}
	&(p_y+q_\mrm{e}A_y)(t=-\infty) = 0\\
	&(p_z+q_\mrm{e}A_z)(t=-\infty) = 0\mrm{.}
\end{align}
{everywhere} in the region where the later interaction takes place.
Thus, Eqs.~(\ref{eq:p_y_2}),~(\ref{eq:p_z_2}) dictate that here for all times
\begin{align}
	&p_y+q_\mrm{e}A_y = 0\\
	&p_z+q_\mrm{e}A_z = 0\mrm{.}
\end{align}
Rewriting this equation again in terms of the electric field, we obtain Eq.~(1) of the main article
\begin{equation}
\partial_t\pvec_\perp=q_\mrm{e}\Evec_\perp\,\mrm{,}\label{eq:material} 
\end{equation}
with $\pvec_\perp=(p_y,p_z)^T$ and $\Evec_\perp=(E_y,E_z)^T$. This equation needs to be coupled to Maxwell's equations
\begin{align}
	\partial_x E_y &= -\partial_t B_z \quad
	&-\partial_x B_z &= \frac{1}{c^2}E_y+\frac{\mu_0q_\mrm{e}}{m_\mrm{e}}\frac{n_\mrm{e}}{\gamma_\mrm{e}}p_y\\
	\partial_x E_z &= \partial_t B_y \qquad
	&\partial_x B_y &= \frac{1}{c^2}E_z+\frac{\mu_0q_\mrm{e}}{m_\mrm{e}}\frac{n_\mrm{e}}{\gamma_\mrm{e}}p_z\mrm{,}\label{eq:Maxwell}
\end{align}
where we have used that $\Jvec=q_\mrm{e}n_\mrm{e}\pvec/(m_\mrm{e}\gamma_\mrm{e})$, $n_\mrm{e}$ is the electron beam density, $m_\mrm{e}$ is the electron mass and $\gamma_\mrm{e}$ is the gamma factor.
This model [Eqs.~(\ref{eq:material})-(\ref{eq:Maxwell})] describes the EBDA for a given space-time evolution of $n_\mrm{e}/\gamma_\mrm{e}$. To obtain the result in Fig.~4b of the main article, we solved this model assuming that the electron beam moves with a constant speed and has an unperturbed Gaussian density profile.
For this purpose, the finite-difference-time-domain Yee-scheme solver ARCTIC~\cite{Taflove1995} was used. 
This approach has been successfully benchmarked against particle-in-cell simulations~(see Fig.~4(b) of the main article). 

\begin{center}
	\begin{figure*}[t]
		\centering
		\includegraphics[width=2.\columnwidth]{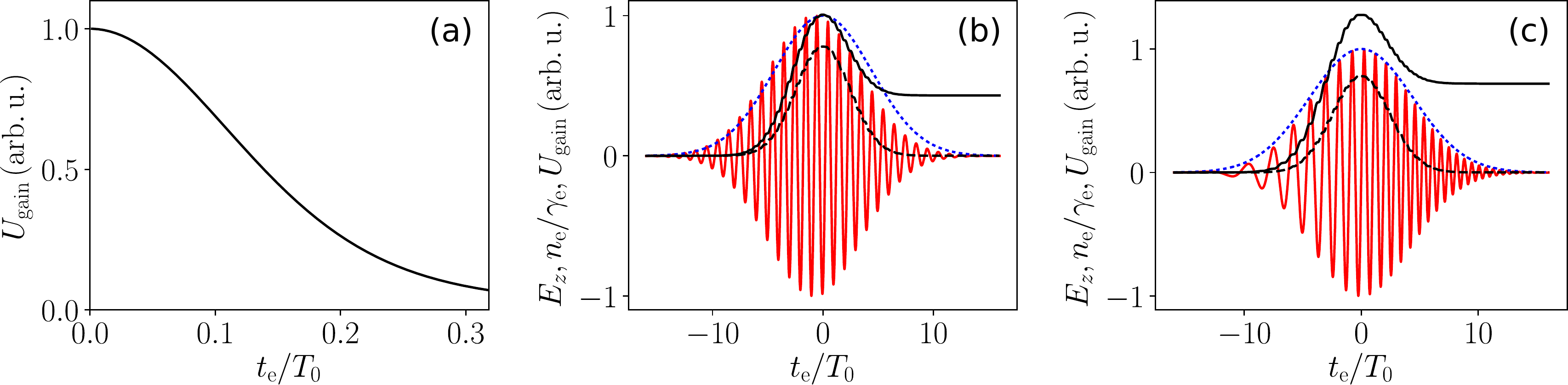}
		\caption{(a)~Final energy density gain versus electron beam duration computed from Eq.~(\ref{eq:U_gain}) assuming an unperturbed reflected seed pulse electric field, a constant charge and the parameter $t_0=0.21T_0$ and $t_\mrm{d}=0.05T_0$ same as in Fig.~4(a) of the main article.
			(b)~Visualization of the amplification mechanism for {different delays of the electron beam}: electric field approximated by the reflected seed pulse field~(solid red line), electron density~(dotted blue line), time-dependent energy density gain $U_\mrm{gain}$ with delay $t_\mrm{d}=0.8T_0$~(solid black line) and without delay~(dashed black line). 
			(c)~Visualization of the amplification mechanism for a long electron beam and chirped seed pulse: electric field approximated by the unperturbed chirped reflected seed pulse field with $C=0.2T_0^{-2}$~(solid red line), electron density~(dotted blue line), time-dependent energy density gain $U_\mrm{gain}$ with chirp~(solid black line) and without chirp~(dashed black line).
			In (b) and (c) $t_\mrm{e}=6.4T_0$ and $t_0=5.9T_0$.}
		\label{fig:U_gain}
	\end{figure*}
\end{center}

In the following, we use Eq.~(\ref{eq:material}) and  Poynting's theorem to show when the seed pulse can be amplified by the electron beam.  Poynting's theorem reads
\begin{equation}
\nabla\cdot\Svec+\Jvec\cdot\Evec+\partial_t u = 0\mrm{,}
\end{equation}
where $u$ is the electromagnetic energy density and $\Svec=\Evec\times\Bvec/\mu_0$ is the Poynting vector~\cite{Jackson}. Using $u({x,}\pm\infty)=0$, we obtain the energy density gain
\begin{equation}
U_\mrm{gain}(x)=\int\limits_{-\infty}^{+\infty}\nabla\cdot\Svec\,d\tau=-\int\limits_{-\infty}^{+\infty}\left(J_xE_x+\Jvec_\perp\cdot\Evec_\perp\right)\,d\tau
\end{equation}
at one particular spatial position. The first term on the right-hand side can be rewritten using Eq.~(\ref{eq:p_x}) and $J_x=q_\mrm{e}n_\mrm{e}p_x/(m_\mrm{e}\gamma_\mrm{e})$ as
\begin{align}
	\int\limits_{-\infty}^{+\infty} J_xE_x\,d\tau=\int\limits_{-\infty}^{+\infty} q_\mrm{e}\frac{n_\mrm{e}}{\gamma_\mrm{e}}\Big[&\frac{1}{2}\left(\partial_t p_x^2 + v_x \partial_x p_x^2\right) \nonumber\\ &- p_x\left(v_y B_z - v_z B_y\right)\Big]\,d\tau\,\mrm{.}\label{eq:long}
\end{align}
We assume that the electron beam moves uniformly along $x$ with the speed $v_x=v_\mrm{b}$ and thus $p_x^2$ is a function of $x-v_\mrm{b} t$. This makes the terms in the first round bracket on the right-hand side of Eq.~(\ref{eq:long}) vanish identically. The second term can be approximated as
\begin{align}
	-p_x\left(v_y B_z - v_z B_y\right) \approx c\left(-p_y B_z + p_z B_y\right)\approx p_y E_y + p_z E_z
\end{align}
using $v_x \approx c$ and $E_z \approx c B_y$ as well as $E_y \approx -c B_z$ for a forward propagating electromagnetic wave and thus
\begin{align}
	\int\limits_{-\infty}^{+\infty} J_xE_x\,d\tau\approx\int\limits_{-\infty}^{+\infty}\Jvec_\perp\cdot\Evec_\perp\,d\tau\,\mrm{.}
\end{align}
We now focus on the time-integral of $\Jvec_\perp\cdot\Evec_\perp$.
Using Eq.~(\ref{eq:material}) and $\Jvec_\perp=q_\mrm{e}n_\mrm{e}\pvec_\perp/(m_\mrm{e}\gamma_\mrm{e})$ we deduce
\begin{align}
	U_\mrm{gain}(x,\infty)&= -\frac{1}{m_\mrm{e}}\int\limits_{-\infty}^\infty\partial_\tau\left|\pvec_\perp\right|^2\frac{n_\mrm{e}}{\gamma_\mrm{e}}\,d\tau\\
	&=-\frac{\left|\pvec_\perp\right|^2}{m_\mrm{e}}\frac{n_\mrm{e}}{\gamma_\mrm{e}}\Bigg|_{t=-\infty}^{t=+\infty}+\int\limits_{-\infty}^\infty\frac{\left|\pvec_\perp\right|^2}{2m_\mrm{e}}\partial_\tau\left(\frac{n_\mrm{e}}{\gamma_\mrm{e}}\right)\,d\tau\nonumber\\
	&=\frac{1}{m_\mrm{e}}\int\limits_{-\infty}^\infty\left|\pvec_\perp\right|^2\partial_\tau\left(\frac{n_\mrm{e}}{\gamma_\mrm{e}}\right)\,d\tau\,\mrm{.}
\end{align}
Finally, we generalize the definition of the energy density gain $U_\mrm{gain}$ and using Eqs.~(\ref{eq:Gauge}),~(\ref{eq:material}) we obtain for all $t$
\begin{equation}
U_\mrm{gain}(x,t) = \frac{q_\mrm{e}^2}{m_\mrm{e}}\int\limits_{-\infty}^t\left|\Avec_\perp\right|^2\partial_\tau\left(\frac{n_\mrm{e}}{\gamma_\mrm{e}}\right)\,d\tau\,\mrm{.}\label{eq:U_gain}
\end{equation}
As can be seen from this equation, a time-increasing value of $n_\mrm{e}/\gamma_\mrm{e}$ contributes to an energy gain, while a decrease in time induces a loss.

To quantify the energy density gain using Eq.~(\ref{eq:U_gain}) it is necessary to know the vector potential $\Avec_\perp$ or the electric field $\Evec_\perp=-\partial_t\Avec_\perp$. This is in general only possible by solving Eqs.~(\ref{eq:material})-(\ref{eq:Maxwell}) as has been done for Fig.~4b of the main article. 
To obtain an estimation of the energy density gain without solving the full problem, the electric field can be approximated as the unperturbed seed electric field~(see discussion of Fig.~4(a) of the main article). 

This simplified approach can be used to give predictions about the energy gain and the amplification process.  First, we consider the energy density gain for different electron beam duration keeping the total charge, i.e. the time-integral of $n_\mrm{e}$, constant. An example with a single-cycle seed pulse following the discussion of Fig.~4 in the main article is shown in Fig.~\ref{fig:U_gain}(a). As predicted by Eq.~(6) of the main article, the gain {goes to a} constant {value} for $t_\mrm{e}/T_0\rightarrow0$ and decreases for longer electron bunches. 

Second, one may ask whether it is possible to amplify many-cycle seed pulses with electron bunches that are longer than the seed wavelength. In the main article we have shown an example in Fig.~5, that amplification is possible for electron beam duration shorter than the seed wavelength. For longer electron beam durations, according to Eq.~(\ref{eq:U_gain}), the amplification at the rising edge of $n_\mrm{e}/\gamma_\mrm{e}$ is possible independently of the evolution of $\Avec_\perp$ and thus also for many-cycle pulses.
If $t_\mrm{d}=0$, then energy is gained at the rising edge of $n_\mrm{e}/\gamma_\mrm{e}$ and completely released at its descending edge. Thus no final energy gain is expected~[see Fig.~\ref{fig:U_gain}(b), dashed black line]. However, an asymmetry of the seed pulse with respect to the electron beam can introduce a final gain: For example if the electron beam is delayed with respect to the seed pulse~[see Fig.~\ref{fig:U_gain}(b), solid black line], or if the seed pulse has been already deformed due to loss at the one edge and gain at the other edge, or {in the vicinity of} the standing mirror when, in addition to the reflected, the incoming part of the seed pulse interacts with the electron bunch in {counter}-propagation. Moreover, the asymmetry can be introduced actively using asymmetric electron bunches or a positively chirped seed pulse with the electric field 
\begin{equation}
E_z(t) =  E_z^\mrm{max}\sin\left(\omega_0t+Ct^2\right)\exp\left(-t^2/t_0^2\right)\,\mrm{.}
\end{equation}
as presented in Fig.~\ref{fig:U_gain}(c). In contrast to the case without chirp~(dashed black line), one obtains a final energy gain even for $t_\mrm{d}=0$. However, one should keep in mind that Eq.~(\ref{eq:U_gain}) using the unperturbed reflected seed electric field can give only a {qualitative} prediction and should be confirmed by Maxwell-consistent modeling accounting for the modifications of the electric field in the future.

\section{Electron-beam-driven amplification vs transition radiation}
\label{app:TR}
\begin{figure}
	\centering
	\includegraphics[width=1.0\columnwidth]{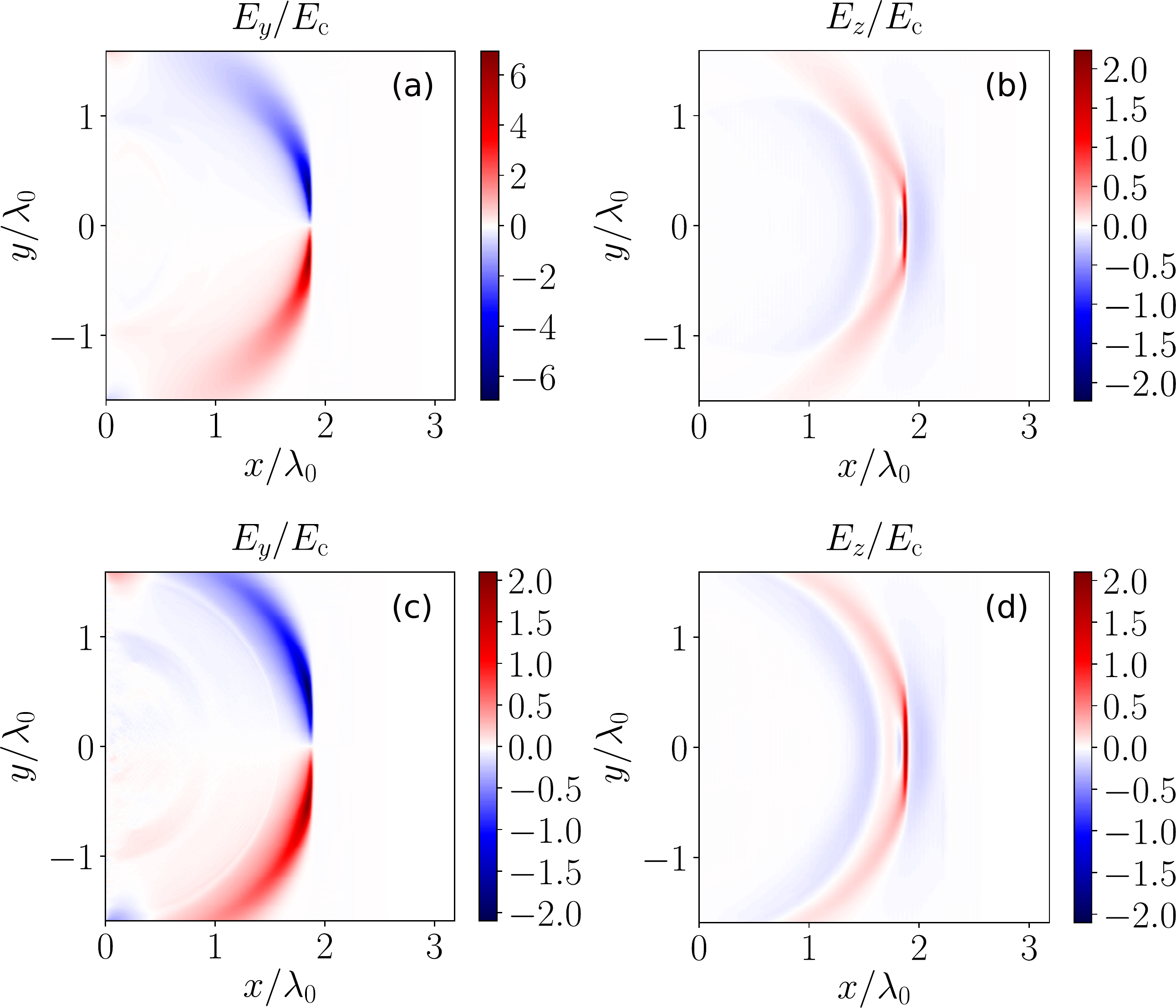}
	\caption{Electric fields snapshot after the interaction of a strongly focused low-frequency pulse with an electron beam passing the standing mirror at $x=0$: (a, c)~The field $E_y$ associated with the transition~(TR) radiation and the field $E_z$ associated with the radiation from the electron-beam-driven amplification~(EBDA). In (a, b) we selected as in the main article $\gamma_\mrm{e}=20$ and in (c, d) a less energetic electron beam with $\gamma_\mrm{e}=7$. Other parameters are as in the main article: $t_\mrm{e}=0.016T_0$, $y_0=0.3\lambda_0$, $n_\mrm{e}^\mrm{max}=28.3n_\mrm{c}$, $\gamma_\mrm{e}=20$, $t_0=0.21T_0$ and $E_0^\mrm{in}=E_\mrm{c}$, where $E_\mrm{c}=cm_\mrm{e}\omega_0/q_\mrm{e}$. The simulation was performed with 1600 points per \EP\ carrier wavelength along $x$, 100 points along $y$ and 1608 points per \EP\ carrier oscillation. For every species 100 particles per cell were used. The mirror has been modeled as a dense electron-proton plasma with thickness $1.4c/\omega_0$.}
	\label{fig:2D_sup}
\end{figure}

When a relativistically fast electron beam passes through a conducting foil, then even without a seed pulse a short electromagnetic pulse is created. This radiation is called transition radiation~(TR)~\cite{Jackson,Schroeder2004}.
{One distinguishes between incoherent transition radiation~(ITR) which scales with $n_\mrm{e}$ and the typically much stronger coherent transition radiation~(CTR) which scales with $n_\mrm{e}^2$. 
A signature of TR can be seen in Fig.~\ref{fig:2D_sup}(a,c).} Since the current emitting TR is longitudinal, TR is radially polarized in 3D having a doughnut-shaped transverse radiation profile with a sharp zero in its center. The duration of the TR pulse is determined by the gamma-factor of the electrons $\gamma_\mrm{e}$. The larger $\gamma_\mrm{e}$, the shorter the pulse{. However, the pulse duration is typically longer than the duration of the electron bunch and is limited by its transverse dimensions resulting in ps-long sub-cycle pulses in the THz frequency range~\cite{Leemans2004,Wu2013,Liao2016}.}

It is important to distinguish TR from the electron-beam-driven amplification~(EBDA) which is the central subject of the main article. One should note that no TR exists in a 1D system with translational invariance in $y$ and $z$. In 3D, TR is radially polarized with a doughnut-shaped transverse profile while the EBDA radiation is linearly polarized with a Gaussian transverse profile for a linearly polarized Gaussian seed pulse. In addition, the TR pulse {which is dominated by CTR is} longer than the \EB~duration~{\cite{Schroeder2004}} while the EBDA pulse is typically as long as the \EB. Due to its Gaussian shape and shorter pulse duration which corresponds to a higher central frequency, the EBDA pulse is more collimated than the TR and can be separated in the far-field. This also implies that the EBDA pulse can be focused more strongly than the TR pulse. Moreover for comparison of the two mechanisms one can use that the TR beam has a sharp zero in its center, where the EBDA beam has its largest field strength.

For numerical simulations, the 2D set-up is advantageous because it allows us to separate the EBDA radiation from the TR radiation already in the near-field if the linearly polarized seed pulse is chosen to be $E_z$-polarized, where $z$ is the translation invariant direction. The TR is generated by the $J_x$ current, where $x$ is the \EB ~propagation axis. Thus, it generates TR with the field components $E_x$, $E_y$ and $B_z$. The $E_z$-polarized seed pulse drives radiation only with field components $B_x$, $B_y$ and $E_z$. 

Transition radiation is quite intense for the \EB~parameters considered in the main article. As can be seen in Figs.~\ref{fig:2D_sup}(a,b), for this example it is more intense than the EBDA pulse. 
{However, the relative intensity of TR and EBDA depends on both the seed pulse amplitude and electron beam energy. 
On one hand, the field strength of the TR is larger than for EBDA radiation in case of weak seed pulses but less intense for relativistic seed pulses as can be seen in Fig.~\ref{fig:TR_vs_EBDA}.
On the other hand, TR becomes weaker for less energetic beams, while the radiation from EBDA scales (in the ideal case) with $n_\mrm{e}/\gamma_\mrm{e}$~(see Sec.~\ref{app:EBDA}). Despite the fact that electron beam depletion and propagation effects reduce the EBDA for low \EB~energies, we show in Fig.~\ref{fig:2D_sup}(c,d) that the EBDA radiation can compete with TR even for weaker seed pulses and low beam energies.}

\begin{figure}
	\centering
	\includegraphics[width=0.85\columnwidth]{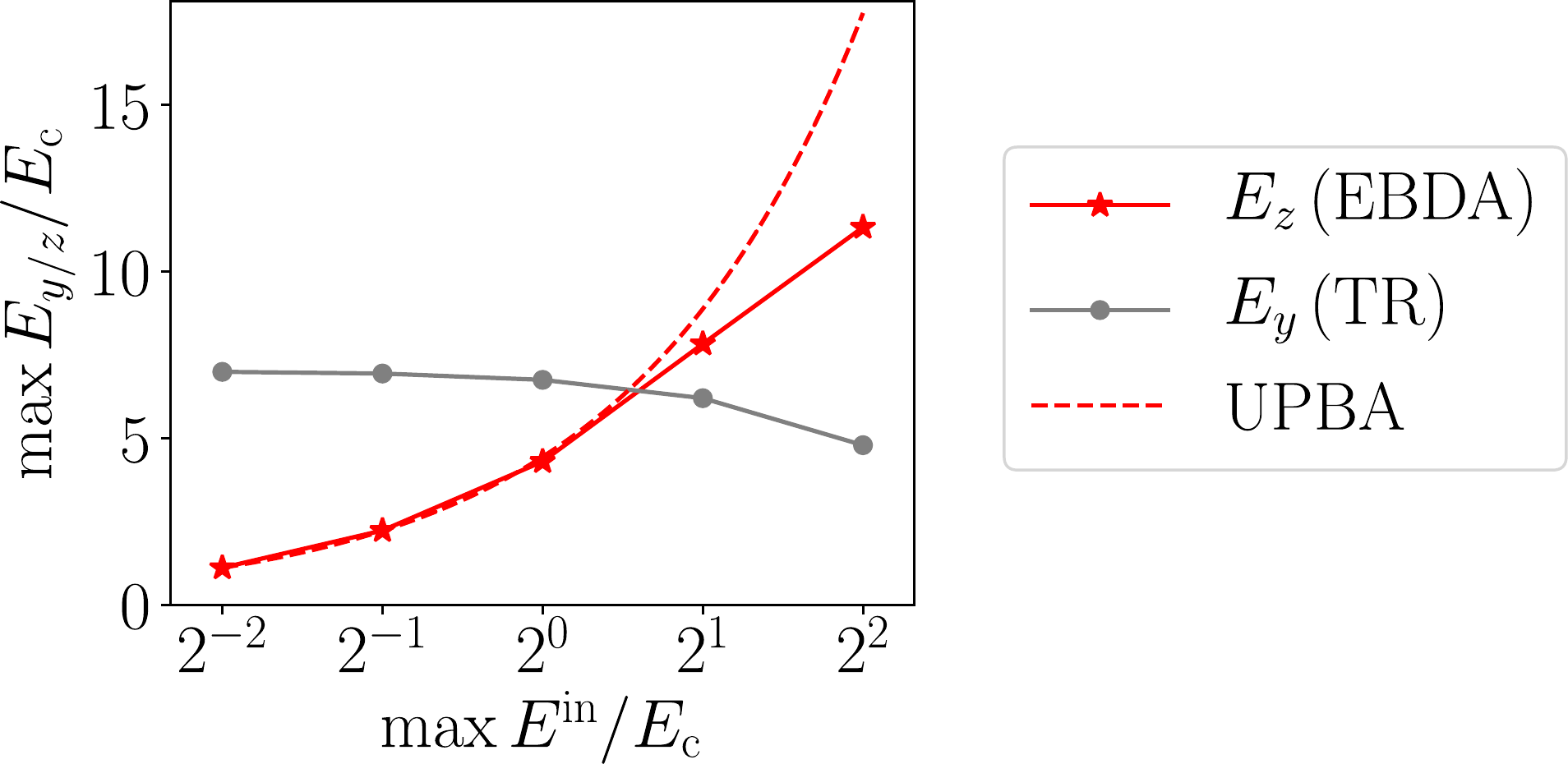}
	\caption{{Comparison of the peak electric fields from the electron-beam-driven amplification~(EBDA) and transition radiation~(TR) for the parameters of Fig.~2 of the main article. The dashed line specifies the expected scaling according to the undepleted pump beam approximation~(UPBA).}}
	\label{fig:TR_vs_EBDA}
\end{figure}

\end{appendices}

\bibliography{mybibfile}

\end{document}